\documentclass[12pt]{article}
\parskip=\medskipamount                 % space between paragraphs
\thicklines                     % thick straight lines for pictures
\usepackage{amsfonts,amssymb,amsmath,amscd}
\usepackage{bbm}
\usepackage{euscript,eufrak}
\usepackage[ps,dvips,matrix,arrow,frame,import,curve,color]{xy}

\newcommand{\bul}{\bullet}
\newcommand{\Tr}{{\rm Tr}}

\newcommand{\RR}{{\mathbb R}}
\newcommand{\CC}{{\mathbb C}}

\newcommand{\cO}{{\mathcal O}}

\newcommand{\tlambda}{{\tilde\lambda}}

\newcommand{\frg}{{\mathfrak g}}

\newcommand{\vol}{{\rm vol}}

\newcommand{\LG}{{{}^L G}}
\newcommand{\ra}{\rightarrow}

\newcommand{\bsigma}{{\bar\sigma}}

\newcommand{\tP}{{\tilde P}}
\newcommand{\ttau}{{\tilde\tau}}
\newcommand{\trho}{{\tilde\rho}}
\newcommand{\Map}{{\rm Map}}

\begin{document}

\title{A-models in three and four dimensions}

\author{Anton Kapustin, Ketan Vyas \\ {\it \small California
Institute of Technology}}

\begin{titlepage}
\maketitle

\abstract{We introduce and study a new 3d Topological Field Theory which can be associated to any compact real manifold $X$. This TFT is analogous to the 2d A-model and reduces to it upon compactification on an interval with suitable boundary conditions.  It plays a role in 3d mirror symmetry as well as in the physical approach to the geometric Langlands duality. A similar TFT can be defined in four dimensions.}

\end{titlepage}

\section{Introduction}

The two most well-known topological field theories (TFTs) in two dimensions are the A and B models defined in \cite{Witten:mir}. These are topological sigma-models whose target spaces are real symplectic and complex Calabi-Yau manifolds respectively. Two-dimensional mirror symmetry acts by exchanging these two kinds of 2d TFTs. In three dimensions there exists a topological sigma-model analogous to the B-model: the Rozansky-Witten model. Its target space is a complex symplectic manifold. Upon reduction on a circle it reduces to the B-model with the same target. The Rozansky-Witten model has been studied in \cite{RW,KRS,KR}. In this paper we construct a 3d analogue of the A-model. Its bosonic fields are a map $\phi:M\ra X$, where $M$ is the 3d worldvolume, and $X$ is an arbitrary real manifold, and a field $\tau\in T^*M\otimes \phi^*TX$. It reduces to the A-model with target $T^*X$ upon compactification on an interval with suitable boundary conditions. We discuss topological observables, both local and nonlocal, and construct boundary conditions for this model. In section 9 we define a similar TFT in four dimensions. 

The 3d A-model plays a role in 3d mirror symmetry. Its gauged version also arises in the study of the $N=4$ $d=4$ super-Yang-Mills theory compactified on a circle and therefore plays an important role in the physical approach to the geometric Langlands duality. These applications of the 3d A-model are sketched in section 10.

A.K. would like to thank Lev Rozansky for useful discussions. This work was supported in part by the DOE grant DE-FG02-92ER40701.

\section{Action and BRST transformations of the 3d A-model}

Let $X$ be a real manifold. We pick a Riemannian metric $g_{ij}$ on $X$ and denote by $\Gamma^i_{jk}$ the components of the Levi-Civita connection on $X$. 
The bosonic fields of the model are a map $\phi:M\ra X$ and a 1-form $\tau\in T^*M\otimes \phi^*TX$. The fermionic fields are 0-forms $\eta,\beta\in\phi^*TX$, and 1-forms $\psi,\chi\in T^*M\otimes \phi^*TX$. Their BRST transformations are
\begin{align}
\delta\phi^i &=\eta^i,\\
\delta\eta^i &=0,\\
\delta\tau^i &=\psi^i-\Gamma^i_{jk}\eta^j\tau^k,\label{eq:deltatau}\\
\delta\psi^i &=\frac12 R^i_{klj}\eta^l\eta^j\tau^k-\Gamma^i_{jk}\eta^j\psi^k,\label{eq:deltapsi}\\
\delta\beta^i&=D^\star\tau^i-\Gamma^i_{jk}\eta^j\beta^k=d^\star\tau^i+\Gamma^i_{jk} \langle d\phi^j,\tau^k\rangle-\Gamma^i_{jk}\eta^j\beta^k ,\\
\delta\chi^i&=d\phi^i-\star D\tau^i-\Gamma^i_{jk}\eta^j\chi^k=d\phi^i- \star(d\tau^i+\Gamma^i_{jk}d\phi^j\tau^k)-\Gamma^i_{jk}\eta^j\chi^k.
\end{align}
Here angular brackets denote scalar product with respect to a Riemannian metric on $M$, $\star$ is the 3d Hodge star operator, $d^\star=\star d\star$, and $R^i_{klj}$ are the components of the Riemann curvature tensor.

The BRST transformations obviously satisfy $\delta^2\phi^i=\delta^2\eta^i=0$. Less obviously, they satisfy $\delta^2\tau^i=\delta^2\psi^i=0$. 
While $\delta^2\beta$ and $\delta^2\chi$ do not vanish, we will construct the theory so that they are proportional to fermionic equations of motion, so that the above BRST-transformations are nilpotent on-shell. In fact, at this stage it is more convenient to make them nilpotent off-shell, so we introduce two auxiliary fields: a bosonic 0-form $P$ and a bosonic 1-form $\tilde P$, both with values in $\phi^*TX$. We redefine the BRST transformations of $\beta$ and $\chi$ to read
$$
\delta\beta^i=P^i-\Gamma^i_{jk}\eta^j\beta^k,\quad \delta\chi^i=\tP^i-\Gamma^i_{jk}\eta^j\chi^k
$$
and we define 
$$
\delta P^i=\frac12 R^i_{klj}\eta^l\eta^j\beta^k-\Gamma^i_{jk}\eta^j P^k,\quad \delta\tP^i=\frac12 R^i_{klj}\eta^l\eta^j\chi^k-\Gamma^i_{jk}\eta^j\tP^k.
$$
It is easy to check that now $\delta^2=0$ on all fields. The BRST transformations are also covariant with respect to changes of coordinates on $X$.

The action is BRST-exact and chosen so that the equations of motion for $P$ and $\tP$ give
$$
P^i=D^\star\tau^i,\quad \tP^i=d\phi^i-\star D\tau^i.
$$
A suitable action is
$$
{\tilde S}=-\delta \int_M \left[ g_{ij} \chi^i\wedge \star \left(\tP^j-2\left(d\phi^j-\star D\tau^j\right)\right)+g_{ij}\beta^i\wedge\star \left(P^j-2D^\star\tau^j\right)\right]
$$
Integrating out the auxuliary fields we get
$$
{\tilde S}={\tilde S}_{bose}+{\tilde S}_{fermi},
$$
where
$$
{\tilde S}_{bose}=\int_M \left(g_{ij}\left(d\phi^i \wedge \star d\phi^j + D\tau^i \wedge \star D\tau^j+D^\star \tau^i \wedge\star D^\star\tau^j\right)\right)-\int_M g_{ij} D\tau^i \wedge d\phi^j,
$$
and
\begin{equation*}
	\begin{split}
		{\tilde S}_{fermi} & = \int_{M} \bigg( 2 g_{ij} \chi^{i} \wedge \big( D \psi^{j} - \star D \eta^{j} \big) + \frac{1}{2} g_{ij} R^{j}_{klm} \chi^{i} \wedge \big( \star \chi^{k} \eta^{l} \eta^{m} - 4 \tau^{k} \eta^{l} d \phi^{m} \big) \\
		& \quad \qquad - 2 g_{ij} \beta^{i} \wedge \star D^{\star} \psi^{j} + \frac{1}{2} g_{ij} R^{j}_{klm} \beta^{i} \wedge \star \big( \beta^{k} \eta^{l} \eta^{m} + 4 \langle \tau^{k}, d \phi^{l} \rangle \eta^{m} \bigg)
	\end{split}
\end{equation*}
The last term in the bosonic part of the action is somewhat pathological. For example, it is becomes complex when the theory is analytically continued to Minkowski signature. While it is a total derivative, it cannot be discarded if $M$ has a boundary. For this reason we prefer to cancel it by adding to ${\tilde S}$ a topological term
$$
S_{top}=\int_M g_{ij} D\tau^i \wedge d\phi^j.
$$
We define the action of the theory as a sum of the BRST-exact action ${\tilde S}$ and the topological term $S_{top}$:
$$
S={\tilde S}+S_{top}
$$
It appears that $S_{top}$ depends on the metric on the target space, but this is not really so. To see this , we may rewrite it in terms of the 1-form field $\ttau$ valued in $T^*X$:
$$
\ttau_i=g_{ij} \tau^j.
$$
Then $S_{top}$ takes the form
$$
S_{top}=\int_M D\ttau_i \wedge d\phi^i=\int_M d\ttau_i \wedge d\phi^i
$$
We might have worked with the field $\ttau_i$ instead of $\tau^i$ from the very beginning, but we choose not to do so.  

Apart from BRST-invariance we also have $U(1)$ symmetry (ghost number symmetry) with respect to which $\phi$ and $\tau$ are unchaged, $\eta$ and $\psi$ have charge $1$, and $\beta$ and $\chi$ have charge $-1$. 

\section{Observables and deformations}

Local BRST-invariant operators which are 0-forms on $M$ are functions of $\phi^i$ and $\eta^i$ which are annihilated by $\delta$. Such functions can be thought of as closed differential forms on $X$. BRST-exact local observables are exact forms, so the BRST-cohomology coincides with the de Rham cohomology of $X$. The algebra of local observables is the cohomology ring of $X$. There cannot be either perturbative or nonperturbative quantum corrections to theses results, because the Planck constant enters only as the coefficient of a BRST-exact action. This is consistent with the fact that the model does not admit BPS instantons on a 3-manifold without boundary. Indeed, such an instanton would satisfy the BPS equations
$$
d\phi=\star D\tau,\quad D^\star\tau=0,
$$ 
and have a vanishing action. On the other hand, the bosonic part of the action 
$$
S_{bose}=\int_M \left(g_{ij}d\phi\wedge\star d\phi^j+g_{ij} D\tau^i\wedge \star D\tau^j+g_{ij} D^\star\tau^i\wedge \star D^\star\tau^j\right)
$$
can be zero if and only if $\phi$ is a constant map and $\tau^i$ is a harmonic 1-form on $M$ for all $i$. 

Elements of the BRST-cohomology at ghost numbers $3$ and $2$ have a special meaning. Elements of ghost number $3$ correspond to infinitesimal deformations of the theory which preserve BRST-invariance and ghost-number symmetry. Indeed, by applying descent to such an observable $\cO$  three times, we get a 3-form $\cO^{(3)}$ of ghost number zero satisfying
$$
\delta \cO^{(3)}=d\cO^{(2)}.
$$
An integral of $\cO^{(3)}$ can therefore be added to the action and to a first order defines a BRST-invariant deformation. In the present case, such elements are closed 3-forms on $X$. The corresponding deformation of the action is simply an integral of the pull-back of the 3-form from $X$ to $M$. This is a 3d analogue of the B-field. 

Elements of degree $2$ correspond to continuous symmetries. Indeed, by applying descent to such an observable $\cO$ three times we get a 2-form $\cO^{(2)}$ of ghost number zero and a 3-form $\cO^{(3)}$ of ghost number $-1$ satisfying 
$$
d \cO^{(2)}=\delta\cO^{(3)}.
$$
The Hodge-dual of $\cO^{(2)}$ is therefore a current conserved up to BRST-exact terms. In the present case, such elements are closed 2-forms on $X$. 

The theory also admits BRST-invariant line operators. The most obvious ones are obtained by picking a vector bundle on $X$ with a flat connection and considering the holonomy of the pull-back of this connection via the map $\phi$. We may refer to such line operators as Wilson lines. They exist only if $X$ is not simply-connected. It is likely that there exist more complicated line operators associated with submanifolds of $X$; we will not attempt to construct them here.

\section{The quantum space of states}

Let us consider quantization of the theory on a 3-manifold of the form $\Sigma \times\RR$, where we regard $\RR$ at time. The classical vacua of the theory have constant $\phi$, constant $\tau_3^i$ and harmonic 1-forms $\tau^i$. Thus the space of classical vacua can be identified with the total space of $TX\otimes (H^0(\Sigma)\oplus H^1(\Sigma))$. There is also one fermionic zero mode for each bosonic zero mode, so that the BRST operator becomes the de Rham operator. If we are interested in normalizable states, we have to restrict the de Rham complex to square-integrable forms. Recall that the $L^2$-cohomology of a vector space is concentrated in the top degree and is one-dimensional. Therefore for compact $X$ the $L^2$-cohomology of the total space of any vector bundle $E$ over $X$ is the cohomology of $X$ shifted by $\dim E$ in degree. We conclude that for any $\Sigma$ the space of states is isomorphic to the de Rham cohomology of $X$ shifted by $(1+2g)\dim X$, where $g$ is the genus of $\Sigma$. 
As discussed above, there can be no perturbative or nonperturbative corrections to this result.

In the case $\Sigma=S^2$ axioms of TFT say that the space of states is isomorphic to the space of local operators. In our case we see that the two agree except for a shift of  grading. The reason for this shift is the noncompactness of the space of bosonic zero modes coming from the 1-form $\tau$ (more specifically, from its time-like component). When computing the BRST cohomology of states, we restricted the BRST-complex to wavefunctions which are square-integrable differential forms on $TX$. BRST-invariant local operators are independent of $\tau_3$ and therefore create states which are not square-integrable. We may identify the BRST cohomology of local operators with the de Rham cohomology of $TX$, without any restriction on the behavior at infinity. This cohomology is isomorphic to $H^\bul(X)$, without any shift of grading.

\section{The partition function}

As explained above, for a closed $M$ the path-integral localizes on configurations which are very simple: $\phi:M\ra X$ is constant and $\tau^i$ is a harmonic 1-form on $X$, for all $i$. The moduli space of such configurations is the total space of the vector bundle $TX\otimes H^1(M)$ and is noncompact unless $H^1(M)=0$. Thus the partition function is finite only if $H^1(M)=0$, i.e. if $M$ is a homology 3-sphere. For such a manifold the only fermionic zero modes are those of $\eta$ and $\beta$. The moduli space measure is
$$
\frac{1}{\sqrt  g} d^n\phi\, d^n\eta\, d^n\beta
$$ 
where $n=\dim X$, $g=\det g_{ij}$ and the factor $1/\sqrt g$ was inserted to make the measure invariant with respect to reparameterizations of $X$. When we restrict the action to configurations with $\tau=\chi=\psi=0$ and constant $\phi$, the only remaining term is the four-fermion interaction
$$
\int_M \frac{1}{2} g_{ij} R^j_{klm} \beta^i\beta^k \eta^l\eta^m \vol_M.
$$
Integration over $\beta$ gives a factor $\sqrt g$ times the Pfaffian of $R^j_{klm}\eta^l\eta^m$. Thus the partition function is simply the integral of ${\rm Pf}\, R$ over $M$, i.e. the Euler characteristic of $X$, times the one-loop determinants. Finally, it is easy to see that fermionic and bosonic one-loop determinants cancel.

While on a closed 3-manifold the path-integral is saturated by field configurations with constant $\phi$, this is not necessarily so if we allow $M$ to have boundaries. Indeed, in such a case the action of a BRST-invariant configuration is equal to
$$
\int_{\partial M} g_{ij}\tau^i d\phi^j,
$$
which is not necessarily zero. If it is nonzero, than $\phi$ cannot be constant, and therefore such a configuration is a nontrivial instanton solution of the BPS equations.

\section{Boundary conditions}

The most obvious boundary condition is to require the restriction of $\tau$ to the boundary to vanish and to impose the free boundary condition on $\phi$ and the normal component of $\tau_3$. If the boundary is given by the equation $x^3=0$ and the metric near the boundary is taken to be Euclidean, these boundary conditions read
$$
\tau_1^i=\tau_2^i=0,\quad \partial_3\tau_3^i=\partial_3\phi^i=0.
$$
These conditions on bosons are compatible with BPS equations and therefore are a candidate for a BRST-invariant boundary condition. The conditions on fermions are then uniquely determined: on the boundary we must have
$$
\psi_1^i=\psi_2^i=0,\quad \beta=\chi_3^i=0,
$$
with all other fermions unconstrained. We will call this the N boundary, to indicate that $\phi$ satisfies the Neumann condition. 

A complementary boundary condition is to require $\phi$ to map $\partial M$ to a particular point on $X$, i.e. to impose the Dirichlet boundary condition. BRST invariance uniquely determines the boundary conditions on all other fields.Namely, we must have
$$
\tau_3^i=0,\quad \partial_3\tau_1^i=\partial_3\tau_2^i=0,\quad \eta^i=\psi_3^i=0,\quad \chi_1^i=\chi_2^i=0.
$$
We will call this the D boundary, to indicate that $\phi$ satisfies the Dirichlet condition.

We may also consider boundary conditions intermediate between N and D conditions. Let us pick a closed submanifold $Y\subset X$ and require $\phi$ to map $\partial M$ to $Y$. We also impose the Neumann condition $\partial_3\phi^i=0$ on the components of $\phi$ normal to $Y$. BRST-invariance then uniquely determines the boundary conditions for all other fields. In particular, the components of $\tau_1$ and $\tau_2$ normal to $Y$ and components of $\tau_3$ tangent to $Y$ satisfy the Neumann condition, while the components of $\tau_1$ and $\tau_2$ tangent to $Y$ and components of $\tau_3$ normal to $Y$ satisfy the Dirichlet condition. Thus we get one boundary conditions for each submanifold $Y$ of $X$.

Boundary conditions for the 2d A-model can be deformed by a flat abelian gauge field. Similar possibility exists in 3d: one may add to the action a boundary term of the form
\begin{equation}\label{defB}
i \int_{\partial M} \phi^*B,
\end{equation}
where $B$ is a closed 2-form on the submanifold $Y$. This is in fact the most general deformation possible. To classify boundary deformations systematically, one considers a BRST-invariant boundary observable $\cO$ with ghost number two. A deformation of the action can be obtained by integrating over $\partial M$ the descendant $\cO^{(2)}$, which is a 2-form of ghost number zero satisfying
\begin{equation*}
\delta \cO^{(2)}=d\cO^{(1)},\quad \delta\cO^{(1)}=d\cO.
\end{equation*}
In our case boundary observables are BRST-invariant functions of $\phi$ and $\eta$. Since $\phi$ on the boundary lies in $Y$ and $\eta$ is tangent to $Y$, one may identify the space of boundary observables with closed differential forms on $Y$. Ghost-number two observables are precisely closed 2-forms on $Y$, and the corresponding deformation of the action is of the form (\ref{defB}).

In the 2d case one can consider adding boundary degrees of freedom, leading to flat vector bundles over $Y$ (which is Lagrangian in the 2d case). Similarly, one can consider adding boundary degrees of freedom in the 3d A-model. Such boundary degrees of freedom are described by a 2d TFT ``fibered'' over $Y$. For example, one may take a family of 2d A-models parameterized by points of $Y$. We leave the construction of the corresponding boundary action for future work. 

\section{Boundary line operators}

Boundary conditions in any 3d TFT are objects of a 2-category (see e.g. \cite{KRS,KSV}). 1-morphisms in this 2-category are boundary defect lines separating different boundary conditions, and 2-morphisms are local operators sitting on junctions of boundary defect lines. In particular, boundary line operators on any particular boundary form a monoidal category (i.e. a category with an associative but not necessarily commutative tensor product). Consider for example a boundary condition associated to a submanifold $Y$. Given any vector bundle on $Y$ with a connection $A$ one can define a boundary line operator as the holonomy of a pull-back of $A$. This is a kind of a boundary Wilson line, and it is BRST-invariant if and only if $A$ is flat. On the classical level, fusion of two Wilson lines with connections $A_1$ and $A_2$ gives another Wilson line with a connection $A_1\otimes 1+1\otimes A_2$. Thus the monoidal structure for boundary Wilson lines corresponds to the tensor product of the flat vector bundles. There can be no quantum corrections to this result, either perturbative or nonperturbative. 

To analyze boundary line operators systematically, one can compactify the 3d A-model on an interval. The category of branes in the resulting 2d TFT can be identified with the category of boundary line operators.\footnote{The monoidal structure cannot be determined from the knowledge of the 2d TFT alone.} 

As an example, consider the N condition. This condition sets the components of $\tau$ tangent to the boundary to zero. Thus the only bosonic fields in the effective 2d TFT will be $\phi^i$ and $\tau_3^i=\lambda^i$. The BPS equations reduce to
$$
d\phi^i=*D\lambda^i.
$$
This equation looks very much like a holomorphic instanton equation, suggesting that the effective 2d TFT is an A-model. In fact, one can rewrite the above equation as a condition for a map $\Phi=(\phi,\lambda)$ from the worldsheet $\Sigma$ to $TX$ to be pseudoholomorphic, provided we choose a suitable almost-complex structure on $TX$. This is the almost-complex structure defined by the condition that its $+i$ eigenspace is spanned by tangent vectors of the form
$$
\frac{\partial}{\partial\phi^i}-\Gamma^j_{ki}\lambda^k\frac{\partial}{\partial\lambda^j}+i\frac{\partial}{\partial\lambda^i}
$$
In matrix form the almost-complex structure is
$$
J=\begin{pmatrix} \Gamma\lambda & 1 \\ -1-(\Gamma\lambda)^2 & -\Gamma\lambda\end{pmatrix},
$$
where $\Gamma\lambda$ is a matrix with elements $\Gamma^i_{jk}\lambda^k$. This almost-complex structure is not integrable, in general. 

We conclude that the effective 2d TFT is the A-model with target $TX\simeq T^*X$. This means that the category of boundary line operators on the N boundary is equivalent to the Fukaya-Floer category of $T^*X$. Objects of this category are roughly speaking Lagrangian submanifolds of $T^*X$ equipped with flat vector bundles. Wilson lines considered above correspond to the case when this  Lagrangian submanifold is $X$ itself embedded into $T^*X$ as the zero section.  

For other boundary conditions reduction on an interval is much more subtle. We hope to discuss it elsewhere. 

\section{The gauged 3d A-model}

Suppose now that $X$ admits an action of a compact Lie group $G$. We will now show how to couple the 3d A-model with target $X$ to the A-type 3d gauge theory with gauge group $G$. The latter theory is the dimensional reduction of the four-dimensional Donaldson-Witten theory \cite{Witten:DW}. Its bosonic fields are a gauge field $A$, a scalar field $\zeta$ in the adjoint representation of $G$, and a complex scalar field $\sigma$ (also in the adjoint representation). Its fermionic fields are a pair of 1-forms $\lambda$ and $\tilde\lambda$ and a pair of 0-forms $\rho$ and $\tilde\rho$. The BRST transformations of these fields before coupling to topological matter are
\begin{align}
\delta A &=\lambda,\\
\delta\lambda &=-d_A\sigma,\\
\delta\zeta &= \rho,\\
\delta\rho &=[\sigma,\zeta],\\
\delta\sigma &=0,\\
\delta\bsigma &=\trho,\\
\delta\trho &= [\sigma,\bsigma],\\
\delta\tlambda &=\star F-d_A\zeta,
\end{align}
where $d_A$ is the covariant derivative with respect to $A$ and $\bsigma=-\sigma^\dag$. These BRST transformations satisfy $\delta^2=\delta_g(\sigma)$ modulo fermionic equations of motion, where $\delta_g(\sigma)$ is the gauge transformation with the parameter $\sigma$. To write down an action it is convenient to introduce an auxiliary bosonic 1-form $H$ and redefine
$$
\delta\tlambda = H,\quad \delta H=[\sigma,\lambda].
$$
The action is then chosen so that the equations of motion for $H$ set $H=\star F-d_A\zeta$. A suitable action is
$$
S_{gauge}=-\frac{1}{2e^2} \delta \int_M \Tr\left[\tlambda\wedge \star (H-2(\star F-d_A\zeta))+\lambda\wedge\star d_A\bsigma\right].
$$

The group $G$ is assumed to act by isometries on the target manifold $X$ of the 3d A-model. Infinitesimally this action is described by a vector field $V=V^i(\phi)\partial_i$ on $X$ with values in the dual of the Lie algebra $\frg$ of $G$. By definition, an infinitesimal gauge transformation of $\phi^i$ corresponding to an element $a\in\frg$ is
$$
\delta_g(a)\phi^i=V^i(a)
$$
Gauge transformations of fields taking values in $\phi^*TX$ involve derivatives of $V^i$, for example:
$$
\delta_g(a)\eta^i =\eta^k \nabla_k V^i-\Gamma^i_{jk} V^j(a)\eta^k=\eta^k\partial_k V^i(a).
$$
Gauge-covariant derivatives of fields are defined accordingly; for example
$$
D\phi^i=d\phi^i+V^i(A),\quad D\eta^i=d\eta^i+\Gamma^i_{jk}d\phi^j\eta^k+\eta^k\partial_k V^i(A),
$$
where $A$ is the gauge field.

To couple the 3d A-model to the A-type 3d gauge theory we modify the BRST transformations for matter fields so that $\delta^2=\delta_g(\sigma)$ on all fields. The modified transformations are
\begin{align}
\delta\phi^i &=\eta^i,\\
\delta\eta^i &=V^i(\sigma),\\
\delta\tau^i &=\psi^i-\Gamma^i_{jk}\eta^j\tau^k,\label{eq:deltatau}\\
\delta\psi^i &=\frac12 R^i_{klj}\eta^l\eta^j\tau^k +\tau^k\nabla_k V^i(\sigma)-\Gamma^i_{jk}\eta^j\psi^k,\label{eq:deltapsi}\\
\delta\beta^i&=P^i-\Gamma^i_{jk}\eta^j\beta^k ,\\
\delta P^i &=\frac12 R^i_{klj}\eta^l\eta^j\beta^k-\Gamma^i_{jk}\eta^j P^k+\beta^k\nabla_k V^i(\sigma),\\
\delta\chi^i&=\tP^i-\Gamma^i_{jk}\eta^j\chi^k,\\
\delta\tP^i &=\frac12 R^i_{klj}\eta^l\eta^j\chi^k-\Gamma^i_{jk}\eta^j\tP^k.+\chi^k\nabla_k V^i(\sigma).
\end{align}

The action of the gauged 3d A-model is the sum of $S_{gauge}$, a BRST-exact matter action
$$
{\tilde S}'=-\delta \int_M \left[ g_{ij} \chi^i\wedge \star \left(\tP^j-2\left(D\phi^j-\star D\tau^j\right)\right)+g_{ij}\beta^i\wedge\star \left(P^j-2D^\star\tau^j\right)\right],
$$
and a topological term 
$$
S'_{top}=\int_M d(g_{ij}\tau^i D\phi^j)=\int_M g_{ij} D\tau^i D\phi^j-\int_M g_{ij}\tau^i  V^j(F),
$$
where $F=dA+A\wedge A$. 

\section{The 4d A-model}

In this section we construct the 4d analog of this A-model (see Appendix A for a construction of the 4d A-model with target $\mathbb{R}^{N}$ from the $\mathcal{N} = 2$ linear $\sigma$-model with target $\mathbb{H}^{N}$).  Let $X$ be a Riemannian manifold with metric $g_{ij}$, Levi-Civita connection $\Gamma^{i}_{jk}$, and Riemann curvature $R^{i}_{jkl}$.  The bosonic fields are a map $\phi$ from the ``spacetime" manifold $M$ to $X$ and an antiselfdual 2-form $\tau$ on $M$ valued in the pullback of the tangent bundle $TX$.  
\begin{equation*}
	\begin{aligned}
		\phi & \in \textrm{Map} \big( M, X \big) , \\
		\tau & \in \Gamma \big( \phi^{*} TX \otimes \Omega^{2-} \big) .
	\end{aligned}
\end{equation*}
The fermionic fields are a scalar $\eta$, an antiselfdual 2-form $\psi$, and a 1-form $\chi$ on $M$ valued in the pullback of the tangent bundle $TX$,
\begin{equation*}
	\begin{aligned}
		\eta & \in \Gamma \big( \phi^{*} TX \big) , \\
		\psi & \in \Gamma \big( \phi^{*} TX \otimes \Omega^{2-} \big) , \\
		\chi & \in \Gamma \big( \phi^{*} TX \otimes \Omega^{1} \big) . \\
	\end{aligned}
\end{equation*}
Finally, it will be convenient to include an auxiliary bosonic 1-form $P$ on $M$ valued in the pullback of the tangent bundle $TX$,
\begin{equation*}
	P \in \Gamma \big( \phi^{*} TX \otimes \Omega^{1} \big) . \\
\end{equation*}

The BRST variations are
\begin{equation} \label{eq:4d_A_model_BRST_variations}
	\begin{aligned}
		\delta \phi^{i} & = \eta^{i} , \\
		\delta \eta^{i} & = 0, \\
		\delta \tau^{i} & = \psi^{i} - \Gamma^{i}_{jk} \eta^{j} \tau^{k} , \\
		\delta \psi^{i} & = \frac{1}{2} R^{i}_{jkl} \tau^{j} \eta^{k} \eta^{l} - \Gamma^{i}_{jk} \eta^{j} \psi^{k} , \\
		\delta \chi^{i} & = P^{i} - \Gamma^{i}_{jk} \eta^{j} \chi^{k} , \\
		\delta P^{i} & = \frac{1}{2} R^{i}_{jkl} \chi^{j} \eta^{k} \eta^{l} - \Gamma^{i}_{jk} \eta^{j} P^{k} .
	\end{aligned}
\end{equation}
It is not difficult to verify that the BRST transformations are nilpotent.

The action for the A-model is BRST exact up to a topological term,
\begin{equation} \label{eq:4d_A_model_action_Q_exact}
	\tilde{S}  = - \delta \int_{M} g_{ij} \chi^{i} \wedge \star \Big( P^{j} - 2 \big( d \phi^{j} - \star D \tau^{j} \big) \Big) - 2 \int_{M} g_{ij} d \phi^{i} \wedge D \tau^{j} ,
\end{equation}
where
\begin{equation*}
	D \tau^{i} = d \tau^{i} + \Gamma^{i}_{jk} d \phi^{j} \tau^{k} .
\end{equation*}
Performing the BRST variation and eliminating the auxiliary field $P$, we find that
\begin{equation} \label{eq:4d_A_model_action}
	\begin{split}
		S & = \int_{M} \bigg( g_{ij} d \phi^{i} \wedge \star d \phi^{j} + g_{ij} D \tau^{i} \wedge \star D \tau^{j} - 2 g_{ij} \chi^{i} \wedge \star D \eta^{j} \\
		& \quad \qquad - 2 g_{ij} \chi^{i} \wedge D \psi^{j} + \frac{1}{2} g_{ij} R^{j}_{klm} \chi^{i} \wedge \big( \star \chi^{k} \eta^{l} \eta^{m} + 4 \tau^{k} d \phi^{l} \eta^{m} \big) \bigg) .
	\end{split}
\end{equation}

\section{Concluding remarks}

We would like to conclude with some examples of dualities where the 3d A-model naturally appears.

Three-dimensional mirror symmetry is a conjectural isomorphism between low-energy limits of certain $N=4$ $d=3$ supersymmetric gauge theories with matter \cite{IS,Ooguri}. $N=4$ $d=3$ theories can be twisted into 3d topological field theories, and since after twist the energy scale does not matter, mirror-symmetric gauge theories should yield isomorphic 3d TFTs.

A toy example of 3d mirror symmetry is the particle-vortex duality which identifies the $N=4$ $d=3$ $U(1)$ gauge theory with the theory of a hypermultiplet with target $\RR^3\times S^1$. Either theory admits two different twists which we will call A-twist and B-twist. On the gauge theory side, the A-twist gives a 3d topological gauge theory which is a reduction of the Donaldson-Witten 4d TFT to three dimensions. The B-twist gives a B-type gauge theory whose bosonic fields are a gauge field $A$ and a 1-form $\phi$ which can be combined into a BRST-invariant complex connection $A+i\phi$. On the hypermultiplet side the A-twist gives the 3d A-model with target $X=S^1$, while the B-twist gives the Rozansky-Witten model with target $T^*\CC^*$. Three-dimensional mirror symmetry exchanges A and B twists; in particular, it implies that the B-type 3d gauge theory is isomorphic to the 3d A-model with target $S^1$. 

It is conceivable that other mirror pairs of supersymmetric $N=4$ $d=3$ theories can be twisted into a pair of 3d TFTs one of which is a (gauged) 3d A-model.
 
The gauged 3d A-model has an important application to the Montonen-Olive duality of $N=4$ $d=4$ super-Yang-Mills theories and its mathematical counterpart, the geometric Langlands duality. Recall that $N=4$ $d=4$ SYM theory has a twisted version which has two candidate supercommuting BRST operators $Q$ and $\tilde Q$ \cite{KW}. The most general BRST operator one can consider is
$$
Q_t=Q+t\tilde Q,\quad t\in \CC\cup \{\infty\}.
$$
The GL-twisted theory thus has a complex parameter $t$. Montonen-Olive duality acts by exchanging $G$ and the Langlands-dual group $\LG$ and exchanging $t=1$ and $t=i$ (at vanishing theta-angle) \cite{KW}.

It appears that the GL-twisted $N=4$ super-Yang-Mills theory with gauge group $G$ at $t=1$ compactified on a circle is isomorphic to the gauged 3d A-model with target $G$, where $G$ acts on itself by conjugation. On the other hand, it has been argued in \cite{KSV} that the GL-twisted theory with gauge group $\LG$ at $t=i$ compactified on a circle is isomorphic to the gauged Rozansky-Witten model with target ${T^*} \LG_\CC$ \cite{KSV}. Thus Montonen-Olive duality gives rise to an isomorphism of the gauged 3d A-model with target $G$ and the gauged Rozansky-Witten model with target ${T^*}\LG_\CC$. In particular, the 2-categories of boundary conditions for these two TFTs must be equivalent. This statement should be regarded as a 2-categorical version of the geometric Langlands duality. For this reason it would be of great interest to study boundary conditions for the gauged 3d A-model. In the case when $X=S^1$ and $G=U(1)$, with a trivial action of the latter on the former, this has been done in \cite{KSV}.

\appendix

\section{The 4d A-Model as a twist of $\mathcal{N} = 2$ $d=4$ Linear $\sigma$-model}

In this appendix, we construct the 4d A-model with target manifold $\mathbb{R}^{N}$ by twisting the 4d $\mathcal{N} = 2$ linear $\sigma$-model on $\mathbb{H}^{N}$.  The bosonic field $\phi$ is a map from the Euclidean spacetime, $\mathbb{R}^{4}$, into the flat hyperk\"ahler target manifold $\mathbb{H}^{N}$ (which is isomorphic to $\mathbb{C}^{2N}$ after a choice of complex structure),
\begin{equation*}
	\phi \in \Map \big( \mathbb{R}^{4}, \mathbb{H}^{N} \big) .
\end{equation*}
The fermionic fields $\psi$ and $\overline{\psi}$ are sections of the spin bundle $S_{+}$ and $S_{-}$ on $\mathbb{R}^{4}$ valued in the pullback of the holomorphic tangent bundle $T \mathbb{H}^{N}$ and antiholomorphic tangent bundle $\overline{T \mathbb{H}^{N}}$, respectively,
\begin{align*}
	\psi & \in \Gamma \big( \phi^{*} T \mathbb{H}^{N} \otimes S_{+} \big), \\
	\overline{\psi} & \in \Gamma \big( \phi^{*} \overline{T \mathbb{H}^{N}} \otimes S_{-} \big) .
\end{align*}
The dynamics of the $\sigma$-model are governed by the action
\begin{equation*}
	S = \int_{\mathbb{R}^{4}} d^{4}x \, \Big( \delta_{i \bar{j}} \partial^{\mu} \phi^{i} \partial_{\mu} \overline{\phi}^{\bar{j}} + i \delta_{i \bar{j}} \overline{\psi}^{\bar{j}} \bar{\sigma}^{\mu} \partial_{\mu} \psi^{i} \Big) .
\end{equation*}

The left action of quarternions on $\mathbb{H}^{N}$ corresponds to $SU(2)_{\mathcal{R}}$ $\mathcal{R}$-symmetry, while the right action of quarternions on $\mathbb{H}^{N}$ gives rise to an additional $SU(2)_{X}$ global symmetry.  Let us introduce the following notation to make the $SU(2)_{\mathcal{R}} \times SU(2)_{X}$ action on $\mathbb{H}^{N}$ manifest,
\begin{align*}
	\phi_{1 1'}^{I} & = \phi^{2I - 1}, & \psi_{1'}^{I} & = \psi^{2I - 1}, \\
	\phi_{1 2'}^{I} & = \phi^{2I}, & \psi_{2'}^{I} & = \psi^{2I}, \\
	\phi_{2 1'}^{I} & = - \overline{\phi}^{\overline{2I}}, & \overline{\psi}^{1' I} & = \overline{\psi}^{\overline{2I - 1}}, \\
	\phi_{2 2'}^{I} & = \overline{\phi}^{\overline{2I - 1}}, & \overline{\psi}^{2' I} & = \overline{\psi}^{\overline{2I}} .		
\end{align*}
where $SU(2)_{\mathcal{R}}$ acts on the unprimed index and $SU(2)_{X}$ acts on the primed index.  Using this notation, we can write the action in a form that is manifestly $SU(2)_{\mathcal{R}} \times SU(2)_{X}$ invariant,
\begin{equation} \label{eq:4d_sigma_model_action}
	S = \int_{\mathbb{R}^{4}} d^{4}x \, \bigg( \frac{1}{2} \delta_{I J} \epsilon^{ab} \epsilon^{a' b'} \partial^{\mu} \phi_{a a'}^{I} \partial_{\mu} \phi_{b b'}^{J} + i \delta_{IJ} \overline{\psi}^{a' I} \bar{\sigma}^{\mu} \partial_{\mu} \psi_{a'}^{J} \bigg) .
\end{equation}
It is not difficult to see that the action respects the following supersymmetry transformations,
\begin{equation} \label{eq:4d_sigma_model_susy_variations}
	\begin{aligned}
		\delta \phi_{a a'}^{I} & = \sqrt{2} \xi_{a} \psi_{a'}^{I} + \sqrt{2} \epsilon_{ab} \epsilon_{a' b'} \bar{\xi}^{b} \bar{\psi}^{b' I}, \\
		\delta \psi_{a'}^{I} & = i \sqrt{2} \sigma^{\mu} \overline{\xi}^{a} \partial_{\mu} \phi_{a a'}^{I}, \\
		\delta \overline{\psi}^{a' I} & = i \sqrt{2} \epsilon^{ab} \epsilon^{a' b'} \bar{\sigma}^{\mu} \xi_{a} \partial_{\mu} \phi_{b b'}^{I} .
	\end{aligned}
\end{equation}
With respect to the $SU(2)_{L} \times SU(2)_{R}$ rotational symmetry, $SU(2)_{\mathcal{R}}$ $\mathcal{R}$-symmetry, and $SU(2)_{X}$ symmetry, the fields and supercharges transforms as shown in the tables above.

\begin{table}[h]
	\begin{center}
		\begin{tabular}[c]{|c|c|c|c|c|}
			\hline
			Field & $SU(2)_{L}$ & $SU(2)_{R}$ & $SU(2)_{\mathcal{R}}$ & $SU(2)_{X}$ \\
			\hline
			$\phi_{a a'}^{I}$ & $1$ & $1$ & $2$ & $2$ \\
			$\psi_{\alpha a'}^{I}$ & $2$ & $1$ & $1$ & $2$ \\
			$\overline{\psi}^{\dot{\alpha} a' I}$ & $1$ & $2$ & $1$ & $2$ \\
			\hline
		\end{tabular} ,
	\end{center}
	\caption{Charges of fields in $\mathcal{N} = 2$ linear $\sigma$-model.} \label{tbl:4d_sigma_model_fields}
\end{table}

\begin{table}[h]
	\begin{center}
		\begin{tabular}[c]{|c|c|c|c|c|}
			\hline
			Field & $SU(2)_{L}$ & $SU(2)_{R}$ & $SU(2)_{\mathcal{R}}$ & $SU(2)_{X}$ \\
			\hline
			$Q_{\alpha a}$ & $2$ & $1$ & $2$ & $1$ \\
			$\overline{Q}^{\dot{\alpha} a}$ & $1$ & $2$ & $2$ & $1$ \\
			\hline
		\end{tabular} ,
	\end{center}
	\caption{Charges of $\mathcal{N} = 2$ supercharges.} \label{tbl:4d_N=2_supercharges}
\end{table}

The A-model is constructed by twisting the $SU(2)_{L}$ subgroup of the rotational symmetry of the $\mathcal{N} = 2$ linear $\sigma$-model by the diagonal subgroup of the $SU(2)_{\mathcal{R}} \times SU(2)_{X}$ symmetry (see Table \ref{tbl:4d_sigma_model_fields} for the charges of fields in the $\mathcal{N} = 2$ linear $\sigma$-model). That is, we replace $SU(2)_L$ with $SU(2)_{L'}$ which is the diagonal subgroup of
\begin{equation*}
SU(2)_L \times SU(2)_{\mathcal{R}} \times SU(2)_{X}
\end{equation*}
The field content of the twisted theory is summarized in Table \ref{tbl:4d_A_model_fields}
\begin{table}[h]
	\begin{center}
		\begin{tabular}[c]{|c|c|c|}
			\hline
			Field & $SU(2)_{L'}$ & $SU(2)_{R}$ \\
			\hline
			$\sigma^{I}$ & $1$ & $1$ \\
			$\tau_{\mu \nu}^{I-}$ & $3$ & $1$ \\
			$\eta^{I}$ & $1$ & $1$ \\
			$\psi_{\mu \nu}^{I-}$ & $3$ & $1$ \\
			$\chi_{\mu}^{I}$ & $2$ & $2$ \\
			\hline
		\end{tabular} ,
	\end{center}
	\caption{Fields in the A-model.} \label{tbl:4d_A_model_fields}
\end{table}

The bosonic field $\sigma$ is a map from spacetime $M$ into the Riemannian manifold $\mathbb{R}^{N}$,
\begin{equation*}
	\sigma \in \Map \big( M, \mathbb{R}^{N} \big) .
\end{equation*}
The bosonic field $\tau$ is an antiselfdual 2-form on $M$ valued in the pullback of the tangent bundle $T \mathbb{R}^{N}$,
\begin{equation*}
	\tau \in \Gamma \big( \sigma^{*} T\mathbb{R}^{N} \otimes \Omega^{2-} \big) .
\end{equation*}
The fermionic fields $\eta$, $\psi$, and $\chi$ are a scalar, antiselfdual 2-form, and 1-form on $M$, respectively, valued in the pullback of the tangent bundle $T \mathbb{R}^{N}$,
\begin{equation*}
	\begin{aligned}
		\eta & \in \Gamma \big( \sigma^{*} T\mathbb{R}^{N} \big) , \\
		\psi & \in \Gamma \big( \sigma^{*} T\mathbb{R}^{N} \otimes \Omega^{2-} \big) , \\
		\chi & \in \Gamma \big( \sigma^{*} T\mathbb{R}^{N} \otimes \Omega^{1} \big) . \\
	\end{aligned}
\end{equation*}

Rewriting the action of the $\mathcal{N} = 2$ linear $\sigma$-model (\ref{eq:4d_sigma_model_action}) in terms of the twisted fields we get
\begin{equation} \label{eq:4d_A_model_action}
	\begin{split}
		S & = \int_{\mathbb{R}^{4}} d^{4}x \, \bigg( \frac{1}{4} \delta_{I J} \partial^{\mu} \sigma^{I} \partial_{\mu} \sigma^{J} + \delta_{I J} \partial^{\nu} \tau_{\mu \nu}^{I-} \partial_{\lambda} \tau^{\mu \lambda J-} \\
		& \quad \qquad + \frac{i}{2} \delta_{IJ} \chi_{\mu}^{I} \partial^{\mu} \eta^{J} - i \delta_{IJ} \chi_{\mu} \partial_{\nu} \psi^{\mu \nu J -} \bigg) .
	\end{split}
\end{equation}
The BRST charge is
\begin{equation*}
	Q_{A} = \epsilon^{\alpha a} Q_{\alpha a}
\end{equation*}
which is a scalar after twisting (see Table \ref{tbl:4d_N=2_supercharges} for the charges of the $\mathcal{N} = 2$ supercharges).  The BRST variations follow from the corresponding supersymmetry transformations (\ref{eq:4d_sigma_model_susy_variations}),
\begin{equation*} 
	\begin{aligned}
		\delta \sigma^{I} & =  \eta^{I} , \\
		\delta \tau_{\mu \nu}^{I-} & = \psi_{\mu \nu}^{I-} , \\
		\delta \eta^{I} & = 0, \\
		\delta \psi_{\mu \nu}^{I-} & = 0, \\
		\delta \chi_{\mu}^{I} & = i \partial_{\mu} \sigma^{I} - 2 i \partial^{\nu} \tau_{\mu \nu}^{I-} .
	\end{aligned}
\end{equation*}
It is easy to see that the BRST transformations and the action are a special case of the BRST transformations and the action of the 4d A-model of section 9, with the auxiliary 1-form $P$ integrated out.


\begin{thebibliography}{99}

\bibitem{Ooguri} J.~de Boer, K.~Hori, H.~Ooguri, Y.~Oz and Z.~Yin, ``Mirror symmetry in three-dimensional gauge theories, SL(2,Z) and  D-brane moduli spaces,''
  Nucl.\ Phys.\  B {\bf 493}, 148 (1997)
  [arXiv:hep-th/9612131].

\bibitem{IS} K.~A.~Intriligator and N.~Seiberg, ``Mirror symmetry in three dimensional gauge theories,''
 Phys.\ Lett.\  B {\bf 387}, 513 (1996) [arXiv:hep-th/9607207].

\bibitem{KR} A.~Kapustin and L.~Rozansky, ``Three-dimensional topological field theory and symplectic algebraic geometry II,''  arXiv:0909.3643 [math.AG].

\bibitem{KRS} A.~Kapustin, L.~Rozansky and N.~Saulina, ``Three-dimensional topological field theory and symplectic algebraic  geometry I,''  Nucl.\ Phys.\  B {\bf 816}, 295 (2009) [arXiv:0810.5415 [hep-th]].

\bibitem{KSV} A.~Kapustin, K.~Setter and K.~Vyas, ``Surface operators in four-dimensional topological gauge theory and Langlands duality,''
  arXiv:1002.0385 [hep-th].

\bibitem{KW} A.~Kapustin and E.~Witten, ``Electric-magnetic duality and the geometric Langlands program,''  Commun. Number Theory Phys. {\bf 1}, 1 (2007).

\bibitem{RW} L.~Rozansky and E.~Witten, ``Hyper-Kaehler geometry and invariants of three-manifolds,'' Selecta Math.\  {\bf 3}, 401 (1997)  [arXiv:hep-th/9612216].

\bibitem{Witten:DW} E.~Witten, ``Topological Quantum Field Theory,''  Commun.\ Math.\ Phys.\  {\bf 117}, 353 (1988).

\bibitem{Witten:mir} E.~Witten, ``Mirror manifolds and topological field theory,'' arXiv:hep-th/9112056.



\end{thebibliography}
\end{document}